\documentclass[12pt]{article}
\usepackage{amsmath}
\usepackage[dvips]{graphicx}
\usepackage{color}
\input amssym

\begin{document}

\def\hd{{^\star}}
\def\prg#1{\medskip\noindent{\bf #1}}  \def\ra{\rightarrow}
\def\lra{\leftrightarrow}              \def\Ra{\Rightarrow}
\def\nin{\noindent}                    \def\pd{\partial}
\def\dis{\displaystyle}                \def\inn{\hook}
\def\grl{{GR$_\Lambda$}}               \def\Lra{{\Leftrightarrow}}
\def\cs{{\scriptstyle\rm CS}}          \def\ads3{{\rm AdS$_3$}}
\def\Leff{\hbox{$\mit\L_{\hspace{.6pt}\rm eff}\,$}}
\def\bull{\raise.25ex\hbox{\vrule height.8ex width.8ex}}
\def\ric{{Ric}}                      \def\tric{{(\widetilde{Ric})}}
\def\tmgl{\hbox{TMG$_\Lambda$}}
\def\Lie{{\cal L}\hspace{-.7em}\raise.25ex\hbox{--}\hspace{.2em}}
\def\sS{\hspace{2pt}S\hspace{-0.83em}\diagup}   \def\hd{{^\star}}
\def\dis{\displaystyle}                 \def\ul#1{\underline{#1}}
\def\mb#1{\hbox{{\boldmath $#1$}}}

\def\hook{\hbox{\vrule height0pt width4pt depth0.3pt
\vrule height7pt width0.3pt depth0.3pt
\vrule height0pt width2pt depth0pt}\hspace{0.8pt}}
\def\semidirect{\;{\rlap{$\supset$}\times}\;}
\def\first{\rm (1ST)}       \def\second{\hspace{-1cm}\rm (2ND)}
\def\bm#1{\hbox{{\boldmath $#1$}}}
\def\nb#1{\marginpar{{\large\bf #1}}}

\def\G{\Gamma}        \def\S{\Sigma}        \def\L{{\mit\Lambda}}
\def\D{\Delta}        \def\Th{\Theta}
\def\a{\alpha}        \def\b{\beta}         \def\g{\gamma}
\def\d{\delta}        \def\m{\mu}           \def\n{\nu}
\def\th{\theta}       \def\k{\kappa}        \def\l{\lambda}
\def\vphi{\varphi}    \def\ve{\varepsilon}  \def\p{\pi}
\def\r{\rho}          \def\Om{\Omega}       \def\om{\omega}
\def\s{\sigma}        \def\t{\tau}          \def\eps{\epsilon}
\def\nab{\nabla}      \def\btz{{\rm BTZ}}   \def\heps{\hat\eps}
\def\bu{{\bar u}}     \def\bv{{\bar v}}     \def\bs{{\bar s}}
\def\bx{{\bar x}}     \def\by{{\bar y}}     \def\bom{{\bar\om}}
\def\tphi{{\tilde\vphi}}  \def\tt{{\tilde t}}

\def\tG{{\tilde G}}   \def\cF{{\cal F}}      \def\bH{{\bar H}}
\def\cL{{\cal L}}     \def\cM{{\cal M }}     \def\cE{{\cal E}}
\def\cH{{\cal H}}     \def\hcH{\hat{\cH}}
\def\cK{{\cal K}}     \def\hcK{\hat{\cK}}    \def\cT{{\cal T}}
\def\cO{{\cal O}}     \def\hcO{\hat{\cal O}} \def\cV{{\cal V}}
\def\tom{{\tilde\omega}}                     \def\cE{{\cal E}}
\def\cR{{\cal R}}    \def\hR{{\hat R}{}}     \def\hL{{\hat\L}}
\def\tb{{\tilde b}}  \def\tA{{\tilde A}}     \def\tv{{\tilde v}}
\def\tT{{\tilde T}}  \def\tR{{\tilde R}}     \def\tcL{{\tilde\cL}}
\def\hy{{\hat y}\hspace{1pt}}  \def\tcO{{\tilde\cO}}

\newcommand{\bc}{\textcolor{blue}}
\newcommand{\gc}{\textcolor{green}}
\newcommand{\cc}{\textcolor{cyan}}
\newcommand{\mc}{\textcolor{magenta}}
\newcommand{\dg}{\textcolor{darkgreen}}
\newcommand{\dc}{\textcolor{darkblue}}


\def\nn{\nonumber}                    \def\vsm{\vspace{-9pt}}
\def\be{\begin{equation}}             \def\ee{\end{equation}}
\def\ba#1{\begin{array}{#1}}          \def\ea{\end{array}}
\def\bea{\begin{eqnarray} }           \def\eea{\end{eqnarray} }
\def\beann{\begin{eqnarray*} }        \def\eeann{\end{eqnarray*} }
\def\beal{\begin{eqalign}}            \def\eeal{\end{eqalign}}
\def\lab#1{\label{eq:#1}}             \def\eq#1{(\ref{eq:#1})}
\def\bsubeq{\begin{subequations}}     \def\esubeq{\end{subequations}}
\def\bitem{\begin{itemize}}           \def\eitem{\end{itemize}}
\renewcommand{\theequation}{\thesection.\arabic{equation}}
\title{Near horizon OTT black hole asymptotic symmetries and soft hair\footnote{This work was partially supported by the Serbian Science Foundation under Grant No. 171031.}}

\author{ B. Cvetkovi\'c and D. Simi\'c\footnote{
        Email addresses: {cbranislav@ipb.ac.rs, dsimic@ipb.ac.rs}} \\
Institute of Physics, University of Belgrade,\\
                     P. O. Box 57, 11001 Belgrade, Serbia}
\date{\today}
\maketitle

\begin{abstract}
We study near horizon geometry of both static and stationary extremal Oliva Tempo Troncoso (OTT) black hole. For each of these black holes,
a set of consistent asymptotic conditions is introduced. The canonical generator for the static configuration is shown to be regular. For the rotating OTT black hole, the asymptotic symmetry is described by the time reparametrization, the chiral Virasoro and centrally extended $u(1)$ Kac-Moody algebra.

\end{abstract}

\section{Introduction}
\setcounter{equation}{0}

The long-standing problem of the origin of black hole entropy is one of the most important opened questions in contemporary physics. There are many proposals  for interpreting the black hole entropy and the corresponding micro-states, such as: entanglement entropy \cite{1},  fuzz-ball \cite{2} or soft hair on the horizon \cite{3,4}. The issue has also been the starting point of many ingenious discoveries, the  most impressive of which is the holographic nature of gravity \cite{5}.

Holographic duality \cite{6} states that gravitational theory in an asymptotically  anti de Sitter (AdS) space-time is dual to a non-gravitational theory defined on conformal boundary of the space-time. Although it still has a status of conjecture, there is a large number of results supporting it. Let us mention, for the purpose of this paper, that  holographic duality offers many insights into the black hole physics, including the black hole information paradox and the origin of the black hole micro-states. Namely, holography provides a derivation of the black hole entropy from the near horizon micro-states via Cardy formula \cite{7}, whose applicability crucially relies on the existence of $2D$ conformal symmetry as a subgroup of the asymptotic symmetry group. In spite of that, the present understanding of the holographic duality is not sufficient for the most general purposes, and we need further generalizations. A notable progress represents a  derivation of the Cardy-like formula  in Warped Conformal Field Theory (WCFT), see ref. \cite{8}.

A particularly interesting generalization is given in \cite{9}, where the authors proposed a hypothesis that the extremal Kerr black hole is dual to the chiral $2D$ CFT. There are indications that this chiral CFT should arise as Discrete Light Cone Quantized  (DLCQ) \cite{10}. More precise, extremal black hole, non necessarily Kerr-like, possesses an intriguing feature that its near horizon geometry is an exact solution of the theory. This allows for the study of physics on the horizon by investigating properties of the near horizon geometry.
For the review of the subject see \cite{z1}.

In this article we analyse the near horizon limit of hairy black hole known as the Oliva-Tempo-Troncoso (OTT) black hole \cite{11}, which is the solution of BHT gravity \cite{12} as well as of the Poincar\'e gauge theory of gravity \cite{13} for the special choice of action parameters. The leading idea of this analysis is study of the influence of hair parameter on micro-states of extremal black hole. Note that obtained near horizon geometries exist, without any reference to extremal black hole, as independent solutions and have their own importance.

Firstly, we analyse the static OTT black hole, which becomes extremal for the specific value of the hair parameter and obtain the corresponding near horizon geometry. Then, we study the asymptotic structure of the near horizon geometry and obtain the asymptotic symmetry group.

Secondly, we continue with studying the rotating  OTT black hole which can be made extremal in two different ways: either  by tuning hair parameter or angular momentum. Extremal one obtained by tuning of the hair parameter, surprisingly, leads to the same near horizon geometry as in the non-rotating case. The extremal OTT black hole with maximal angular momentum leads to the geometry with a richer structure. We conclude, in the end, that asymptotic symmetry is a direct sum of the time reparametrization, the Virasoro algebra and the centrally extended $u(1)$ Kac-Moody algebra. The entropy of the extremal rotating OTT black hole can be expressed in terms of the central extension of the Kac-Moody and the on-shell value of the zero mode Virasoro generator
\be
S=2\pi\sqrt{\frac12L_0^{\it on-shell}\kappa}.
\ee

Our conventions are the same as in Ref. \cite{13}: the Latin indices $(i,
j, k, ...)$ refer to the local Lorentz frame, the Greek indices
$(\m,\n,\r, ...)$ refer to the coordinate frame, $e^i$ is the orthonormal
triad (coframe 1-form), $\om^{ij}$ is the Lorentz connection (1-form), the
respective field strengths are the torsion $T^i=de^i+\om^i{_m}\wedge e^m$
and the curvature $R^{ij}=d\om^{ij}+\om^i{_k}\wedge \om^{kj}$ (2-forms),
the frame $h_i$ dual to $e^j$ is defined by $h_i\inn e^j=\d_i{^j}$, the
signature of the metric is $(+,-,-)$, totally antisymmetric symbol
$\ve^{ijk}$ is normalized to $\ve^{012}=1$, the Lie dual of an
antisymmetric form $X^{ij}$ is $X_i:=-\ve_{ijk}X^{jk}/2$, the Hodge dual
of a form $\a$ is $\hd\a$, and the exterior product of forms is implicit.

\newpage
\section{Conformally flat Riemannian solutions in PGT}\label{sec2}
\setcounter{equation}{0}

In the sector with a unique  AdS ground state, the BHT gravity possesses an interesting black hole solution, the OTT black hole \cite{11}. One of
the key features of this solution is its conformal flatness, which enables it to be also a Riemannian solution of PGT in vacuum \cite{13}, for the special choice of the Lagrangian parameters.

The general parity preserving  Lagrangian 3-form of PGT, which is mostly quadratic in field strengths is given by:
\bea\lab{2.1}
&&L_G=-a_0\ve_{ijk}e^iR^{jk}-\frac13\L_0\ve_{ijk}e^ie^je^k+L_{T^2}+L_{R^2}\,,\nn\\
&&L_{T^2}=T^i\hd\left(a_1{}^{(1)}T_i+a_2{}^{(2)}T_i+a_3{}^{(3)}T_i\right)\,,\nn\\
&&L_{R^2}=\frac12 R^{ij}\hd\left(b_4{}^{(1)}R_{ij}+b_5{}^{(5)}R_{ij}+b_6{}^{(6)}R_{ij}\right)\,.
\eea
where ${}^{(a)}T_i$ and ${}^{(a)}R_{ij}$ are irreducible components of the torsion and the RC curvature, see \cite{14}{,} $a_0=1/{16\pi G}$, $\L_0$ is a cosmological constant, and $(a_1, a_2,a_3)$ and $(b_1,b_2,b_3)$ are the  coupling constants in the torsion and the curvature sector, respectively. In \cite{13}, it was shown that any  conformally flat solution of the BHT gravity (the OTT black hole in particular) is also a Riemannian solution of PGT, provided that
\be
b_4+2b_6=0\,. \lab{2.2}
\ee
The conformal properties of 3D spacetime, where the Weyl curvature
vanishes identically, are characterized by Cotton 2-form $C^i$ \cite{15},   defined by $C^i:=\nab L^i=dL^i+\om^i{_m}L^m$ where $L^m:=\ric^m-\frac{1}{4}Re^m$ is the Schouten
1-form. Conformal flatness of the {space-time} is expressed by the condition  $C^i=0$.

By using the BHT condition that ensures the existence of the
unique maximally symmetric background \cite{13}, the identification \eq{2.2} can be expressed in the following way:
\be
\L_0=-{a_0}/{2\ell^2}\, ,\qquad b_4=2a_0\ell^2\, .              \lab{2.3}
\ee

\section{Canonical generator and conserved charges}
\setcounter{equation}{0}

The usual construction of the canonical generator of the Poincar\'e gauge transformations -- comprising diffeomorphisms  and Lorentz rotations \cite{x1},  makes use of the canonical structure of the theory. The construction can be substantially  simplified by using the first order formulation of the
theory, in which the Lagrangian  (3-form) reads:
$$
L_G=T^i\t_i+\frac{1}{2}R^{ij}\r_{ij}-V(e,\t,\r)\, ,
$$
see  \cite{14}. In this formulation, $\t^m$ and $\r_{ij}$ are independent
dynamical variables, the covariant field momenta conjugate to $e^i$ and
$\om^{ij}$. The presence of the potential $V$ ensures the validity of the on-shell relations
$\t_i=H_i$, $\r_{ij}=H_{ij}$. These relations can be used to transform $L_G$ into its standard
quadratic form \eq{2.1}.

The construction of the  canonical generator $G$ in the first order formulation  can be found in  \cite{14}.
The action of  $G$  on the basic dynamical variables is defined via the
Poisson bracket operation, so that $G$   has to be a differentiable phase space functional. The examination of
the differentiability of $G$  starts from  its variation
\bsubeq
\bea
&&\d G=-\int_\S d^2x(\d G_1+\d G_2)\, ,                         \nn\\
&&\d G_1=\ve^{t\a\b}\xi^\mu\left(e^i{_\mu}\pd_\a\d\t_{i\b}
         +\om^i{_\mu}\pd_\a\d\r_{i\b}+\t^i{_\mu}\pd_\a\d e_{i\b}
         +\r^i{_\mu}\pd_\a\d\om_i{_\b}\right)+\cR\, ,           \\
&&\d G_2=\ve^{t\a\b}\th^i\pd_\a\d\r_{i\b}+\cR\, .\lab{3.1b}
\eea
\esubeq
Here, $\S$ is the spatial section of spacetime, the variation is performed
in the set of adopted asymptotic states, $\cR$ stands for regular
(differentiable) terms, and we use $\r^i$ and $\om^i$, the Lie duals of
$\r_{mn}$ and $\om_{mn}$, to simplify the formulas. Diffeomorphisms are parametrized
by $\xi^\mu$, and the parameters of local Lorentz rotations are $\th^i$.

The explicit form of the Lorentz rotations generator, see \cite{14}, implies that there is only one possible non-regular term on the rhs of the variation of
the Lorentz rotations generator $G_2$, which is of the form \eq{3.1b}.

In general  $\d G\ne\cR$, so that  $G$ is not differentiable. This problem
can be, in principle,  easily resolved  by going over to the improved generator $\tG:=G+\G$,
where the boundary term $\G$ is constructed so that $\d\tG=\cR$.
By making a partial integration in $\d G$, one finds that $\G$ is
defined by the following variational equation
\bsubeq
\bea
&&\d\G=\d \G_1+\d \G_2\,,\nn\\
&&\d \G_1=\int_{\pd\S}\xi^\mu\left(
     e^i{_\m}\d\t_i+\om^i{_\m}\d\r_i
    +\t^i{_\m}\d e_i+\r^i{_\m}\d\om_i\right)\, ,\\
&&\d \G_2=  \int_{\pd\S} \th^i\d\r_i   \,.         \lab{3.2}
\eea
\esubeq

In many cases the asymptotic conditions ensure the regularity of the Lorentz rotations generator and $\G_2=0$. However, it is worth noting that in the particular problem which we are going to solve the contribution to the surface term of the Lorentz rotations generator
is non-trivial, as we shall see in section 5.2.
\section{Static OTT black hole orbifold}
\setcounter{equation}{0}

\prg{Extremal static OTT black hole.}
The metric of the static OTT black hole is given by:
\be
ds^2=N^2dt^2-N^{-2}dr^2-r^2d\vphi^2\,,
\ee
where $N^2=-\m+br+\dis\frac{r^2}{\ell^2}$. Black hole horizons are located at:
$$
r_\pm=\frac{1}2\left(-b\ell^2\pm\ell\sqrt{b^2\ell^2+4\m}\right)\,.
$$
The black hole is {extremal} iff horizons coincide, $r_+=r_-$. This condition is satisfied if $b^2\ell^2+4\m=0$. Let us note that the existence of the extremal black hole horizon implies $b<0$.

\prg{Orbifold. } Let us now consider the following coordinate transformation:
\be
t\ra\frac t{\ve}\,,\qquad r\ra r_++\ve\r\,.
\ee
Now, the metric becomes:
$$
ds^2=\frac{\r^2}{\ell^2}dt^2-\frac{\ell^2}{\r^2}d\r^2-(r_++\ve\r)^2d\vphi^2\,.
$$
In the limit $\ve\ra 0$, the metric (with the prescription $\r\ra r$) reads:
\be
ds^2=\frac{r^2}{\ell^2}dt^2-\frac{\ell^2}{r^2}dr^2-r_+^2d\vphi^2\,.
\ee
It represents a perfectly regular solution, an orbifold.

We choose triad fields in the simple diagonal form:
\bsubeq
\be
e^0=\frac r\ell dt\,,\qquad e^1=\frac{\ell}r dr\,,\qquad e^2=r_+d\vphi\,. \lab{2.4a}
\ee
The Levi-Civita connection that corresponds to the triad field reads
\be
\om^0=0\,,\qquad\om^1=0\,,\qquad\om^2=-\frac{e^0}\ell\,.
\ee
\esubeq
The curvature 2-form has only one non-vanishing component:
\bsubeq
\bea
R^0=0\,,\qquad R^1=0\,,\qquad R^2=\frac1{\ell^2}e^0e^1\,,
\eea
the scalar curvature is constant, $R=\dis\frac 2{\ell^2}$,
and the Ricci and Shoutten one forms are given by:
\bea
&&Ric^0=\frac{e^0}{\ell^2}\,,\qquad Ric^1=\frac{e^1}{\ell^2}\,,\qquad Ric^2=0\,,\nn\\
&&L^0=\frac{e^0}{2\ell^2}\,,\qquad L^1=\frac{e^1}{2\ell^2}\,,\qquad L^2=-\frac{e^2}{2\ell^2}\,.
\eea
\esubeq
The solution is conformally flat (as the OTT black hole), i.e. the Cotton 2-form $C^i=\nab L^i$ vanishes, and it solves equations of motion of both BHT gravity and PGT in the sector $b_4+2b_6=0$.

\subsection{Asymptotic conditions}

Let us consider the following asymptotic conditions for the metric in the region $r\ra\infty$:
\be\lab{4.6}
g_{\m\n}\sim\left(
\ba{ccc}
\cO_{-2}&\cO_2&\cO_1\\
\cO_2&\dis
-\frac{\ell^2}{r^2}+\cO_3&\cO_1\\
\cO_1&\cO_1&\cO_0
\ea
\right)\,,
\ee
where $\cO_n$ denotes a term with asymptotic behaviour $r^{-n}$ or faster. In accordance with \eq{4.6}, the asymptotics of the triad  fields  is given by
\bea\lab{4.7}
e^i{_\m}\sim\left(
\ba{ccc}
\cO_{-1}&\cO_3&\cO_2\\
\cO_1&\dis\frac\ell r+\cO_2&\cO_0\\
\cO_1&\cO_2&\cO_0
\ea
\right)
\eea
The condition $T^i=0$, together with \eq{4.7}, gives the following asymptotics of the spin connection
\bea
\omega^i{_\m}\sim\left(
\ba{ccc}
\cO_1&\cO_2&\cO_1\\
\cO_1&\cO_3&\cO_1\\
\cO_{-1}&\cO_3&\cO_2
\ea
\right)
\eea
The diffeomorphisms that leave the  metric \eq{4.6}  invariant are given by:
\bea
&&\xi^t=T(t)+\cO_3\,,\nn\\
&&\xi^r=r U(\vphi)+\cO_0\,,\nn\\
&&\xi^\vphi=S(\vphi)+\cO_1\,.
\eea

{Lorentz} transformations that leave the asymptotic conditions invariant are
\bea
&&\th^0=\cO_2\,,\qquad
\th^1=\cO_2\nn\\
&&\th^2=\cO_2\,.
\eea

In terms of the Fourier modes $\ell_n:=\d_0(S=e^{in\vphi})$ and $j_n:=\d_0(U=e^{im\vphi})$, the algebra of the residual gauge transformations takes the form of a semi-direct sum of the Virasoro and the Kac-Moody algebra:
\bea
&&[\ell_m,\ell_n]=-i(m-n)\ell_{m+n}\,,\nn\\
&&[\ell_m,j_n]=inj_{m+n}\,,\nn\\
&&[j_n,j_m]=0\,.
\eea

\subsection{Algebra of charges}

The gauge generator is not a priori well-defined because, for given asymptotic conditions, its functional derivatives may be ill-defined, as we already
mentioned in section 3. This problem can be resolved by
construction of the improved generator, defined by adding suitable surface terms \cite{x2}. Since our solution is Riemannian, $\t_i=0$, relation \eq{3.2} reduces to:
\be
\d G=\int_{\pd \S}\xi^\m(\om^i{_\mu}\d\r_i+\r^i{_\m}\d \om_i)
\ee

For the particular asymptotic conditions adopted in this paper, one concludes that the gauge generator is differentiable, so that there is no need for adding any surface term,
\be
\Gamma=0.
\ee
As a consequence, both the central charge of the Virasoro algebra and the level of the $U(1)$ Kac-Moody algebra both vanish.

\section{Near-horizon geometry of rotating OTT}
\setcounter{equation}{0}

\prg{Rotating OTT black hole.} The rotating OTT black hole is defined by the metric
\bsubeq\lab{5.1}
\be
ds^2=N^2dt^2-F^{-2}dr^2-r^2(d\vphi+N_\vphi dt)^2\,,
\ee
where
\bea
&&F=\frac{H}{r}\sqrt{ \frac{H^2}{\ell^2}
   +\frac{b}{2}H\left(1+\eta\right)
   +\frac{b^2\ell^2}{16}\left(1-\eta\right)^2-\mu\eta }\, ,     \nn\\
&&N=AF\, ,\qquad A=1+\dis\frac{b\ell^2}{4H}(1-\eta)\,,          \nn\\
&&N_{\vphi}=\frac{\ell}{2r^2}\sqrt{1-\eta^2}(\m-bH)\, ,         \nn\\
&&H=\sqrt{ r^2-\frac{\m\ell^2}{2}(1-\eta)
          -\frac{b^2\ell^4}{16}\left(1-\eta\right)^2 }\, .
\eea
\esubeq
The roots of $N=0$ are
$$
r_\pm=\ell\sqrt{\frac{1+\eta}{2}}
    \left(-\frac{b\ell}{2}\sqrt{\eta}
          \pm\sqrt{\m+\frac{b^2\ell^2}{4}}\right)\, .
$$
The metric \eq{5.1} depends on three free parameters, $\m$, $b$ and
$\eta$. For $\eta=1$, it represents the static OTT black hole, and for
$b=0$, it reduces to the rotating BTZ black hole with parameters $(m,j)$,
such that $4Gm:=\m$ and $4Gj:=\m\ell\sqrt{1-\eta^2}$.

The  conserved charges of the rotating black hole take the following form:
\bsubeq\lab{5.3}
\bea
&&E=\frac{1}{4G}\left(\m+\frac{1}{4}b^2\ell^2\right)\, ,         \\
&&J=\ell\sqrt{1-\eta^2}\,E\,.
\eea
\esubeq

The rotating OTT black hole is a three-parameter solution, so that the extremal limit can be achieved in two different ways. The first
way is the same as in the non-rotating case, by requiring $4\mu+b^2\ell^2=0$. As a simple consequence, the resulting geometry is the same as if the black hole were non-rotating. This is not a surprising result if we note that, in this case, both energy and angular momentum vanish.

The second way to obtain an extremal black hole is to  take $\eta=0$, which means that angular momentum takes the maximal possible value. This
corresponds to the usual procedure for the Kerr black hole.

The horizon is located at
\be
r_0=\frac{\ell\sqrt{b^2\ell^2+4\mu}}{2\sqrt{2}}.
\ee
The coordinate change is given as
\bea
r\ra r_0+\eps r\\ \nn
t\rightarrow\frac{t}{\eps^2}\\ \nn
\vphi \rightarrow \vphi-\frac{t}{\ell\eps^2}.
\eea
An interesting departure from the usual redefinition of coordinates in the literature is that, in order to obtain a non-singular metric, we have to scale time coordinate with the same parameter used in the rescaling of the radial coordinate to the power of  minus two, instead of the standard minus one.

After changing the coordinates and taking the limit $\eps{\ra}0$, we obtain the near-horizon metric
\be
ds^2=\frac{32(b^2\ell^2+4\mu)}{b^4\ell^4}\frac{r^4}{\ell^4}dt^2-\frac{\ell^2}{r^2}dr^2-r_0^2\left(d\vphi-\frac{16r^2}{b^2\ell^5}dt\right)^2,
\ee
or
\be
ds^2=2r_0^2\frac{16r^2}{b^2\ell^5}dtd\vphi-\frac{\ell^2}{r^2}dr^2-r_0^2d\vphi^2.
\ee
It is convenient to further rescale the time coordinate and obtain a more convenient form of the metric
\be
ds^2=\frac{2r^2r_0}{\ell^2}dtd\vphi-\frac{\ell^2}{r^2}dr^2-r_0^2d\vphi^2\,.
\ee
We {again choose} triad {fields in the diagonal form}
\bea
&&e^0=\frac{r^2}{\ell^2}dt\,,\qquad e^1=\frac{\ell}{r}dr\,,\qquad e^2=\frac{r^2}{\ell^2}dt-r_0d\vphi\,.
\eea
The Levi-Civita connection is given by:
\bea
\om^{01}=-\frac{2e^0}\ell+\frac{e^2}\ell\,,\qquad \om^{02}=\frac{e^1}\ell\,,\qquad \om^{12}=\frac{e^0}\ell\,.
\eea
The solution is maximally {symmetric} and therefore we have:
\bea
R^{ij}=\frac1{\ell^2}e^ie^j\,,\qquad Ric^i=\frac{2e^i}{\ell^2}\,,\qquad L^i=\frac{e^i}{2\ell^2}\,,\qquad C^i=0\,.
\eea

The rotating OTT black hole for $b=0$ reduces to the rotating BTZ black hole. What can be said about the corresponding near-horizon geometries? If we introduce $\r=r^2$, we obtain near-horizon BTZ black hole geometry with two times smaller $\ell$, and a different $r_0$ \cite{z1}. The only trace of the hair parameter is hidden in $r_0$, and it will lead to different values of the central charges. Thus, we are able to recover the results for the near-horizon BTZ black hole geometry from those for OTT black hole, but not in a naive way by simply taking $b=0$.

\subsection{Asymptotic conditions}

We consider the following asymptotic form of the metric
\be\lab{3.6}
g_{\m\n}\sim\left(
\ba{ccc}
\cO_{-1}&\cO_{3}&\cO_{-2}\\
\cO_{3}&\dis
-\frac{\ell^2}{r^2}+\cO_4&\cO_1\\
\cO_{-2}&\cO_1&\cO_0
\ea
\right)\,.
\ee
The {asymptotic form} of the triad  fields is chosen in accordance with the {asymptotic} behaviour of the metric \eq{3.6}
\bea\lab{3.7}
e^i{_\m}\sim\left(
\ba{ccc}
\dis\frac{r^2}{\ell^2}+\cO_{1}&\cO_5&\cO_0\\
\cO_1&\dis\frac\ell r+\cO_3&\cO_0\\
\dis\frac{r^2}{\ell^2}+\cO_1&\cO_5&\cO_0
\ea
\right)
\eea
The asymptotic form of the spin connection reads
\bea\lab{3.8}
\omega^i{_\m}\sim\left(
\ba{ccc}
-\dis\frac{r^2}{\ell^3}+\cO_1&\cO_2&\cO_0\\
\cO_0&-\dis\frac{1}{r}+\cO_2&\cO_0\\
-\dis\frac{r^2}{\ell^3}+\cO_{1}&\cO_2&\cO_0
\ea
\right)
\eea
The condition of vanishing torsion $T^i=0$, together with \eq{3.7} and \eq{3.8}, leads to the following constraints
\bsubeq
\bea
&&\omega^{2}_r+\omega^{0}_r=\mathcal{O}_5\,,\\
&&\omega^{1}_\vphi-\frac{e^1_\vphi}{\ell}=\cO_2\,,\\
&&\frac{e^0_{\ \vphi}}{\ell}-\frac{e^2_{\ \vphi}}{\ell}+\omega^{2}_\vphi -\omega^{0}_\vphi=\cO_2\,,\lab{5.16}\\
&&\omega^{2}_\vphi-\frac{e^2_\vphi}{\ell}=\cO_1\,,\\
&&\omega^{0}_\vphi-\frac{e^0_\vphi}{\ell}=\cO_1\,.
\eea
\esubeq
The diffeomorphisms that leave the  metric \eq{3.6}  invariant are given by
\bea
&&\xi^t=T(t)+\cO_3\,,\nn\\
&&\xi^r=r U(\vphi)+\cO_1\,,\nn\\
&&\xi^\vphi=S(\vphi)+\cO_4\,.
\eea
Lorentz transformations that leave the asymptotic form of the triads and the spin connection invariant are
\bea
&&{\theta^0=\pd_{r}\xi^t\frac{e^2_{\ t}}{e^1_{\ r}}+\cO_2\,,}\nn\\
&&{\theta^1=-\frac{2\xi^r}{r}+\partial_t\xi^t+\cO_4\,,}\nn\\
&&{\theta^2=\frac{e^0_{\ t}}{e^1_{\ r}}\pd_{r}\xi^t+\cO_2\,.}
\eea

\subsection{Algebra of charges}

The improved generator is given by
\be
\tG=G+\G\,.
\ee
A direct calculation yields the surface term
\bea
\G&=&-4a_0\int_{0}^{2\pi}d\vphi\Bigl[T(t)\frac{r^2}{\ell^2}(\omega^0{_\vphi}-\frac{e^0{_\vphi}}{\ell}-\omega^2{_\vphi}+\frac{e^2{_\vphi}}{\ell})\nn\\
&+&S(\vphi)\omega^i{_\vphi} e_{i\vphi}+\left(2U(\vphi)+\pd_{t}T(t)\right)e^1{_\vphi}\Bigr]
\eea
The charge is finite due to the conditions {which follow} from the constraint $T^i=0$.
{By} using the composition law for the local Poincar\'e transformations
\bea
&&\xi^{\prime\prime\mu}=\xi^\a\partial_\a\xi^{\prime\mu}-\xi^{\prime\a}\partial_\a\xi^\mu\,,\nn\\
&&\theta^{\prime\prime i}=\epsilon^i_{\ jk}\theta^j\theta^{\prime k}+\xi^\a\partial_\a\theta^{\prime i}-\xi^{\prime\a}\partial_\a\theta^i
\eea
we derive the Poisson bracket algebra between improved canonical generators (which are also well-defined \cite{y1}).
The Virasoro algebra is not centrally extended
\bea
&&\{L_m,L_n\}=-i(m-n)L_{m+n}\,,\lab{5.20}\\
&&\{L_m,J_n\}=inJ_{m+n}\,,\lab{5.21}
\eea
whereas the Kac-Moody algebra does have a central charge $\kappa$
\be
\{J_m,J_n\}=-i16\pi a_0m\delta_{m+n,0},
\ee
whose value is
\be
\k=16\pi a_0=\frac{\ell}{G}.
\ee
For related works, see \cite{x3,x4}.

The entropy of the extremal OTT black hole $S=\dis\frac{\pi r_0^2}{G}$ can be reproduced in terms of purely algebraic quantities via a peculiar formula
\be
S=2\pi \sqrt{\frac12L^{\it on-shell}_0\kappa},\lab{5.30}
\ee
where $L^{\it on-shell}_0$ is the value of the Virasoro generator $L_0$ on the solution
\be
L^{\it on-shell}_0=\frac{r^2_0}{2\ell G}.
\ee
The entropy formula has a striking resemblance to the entropy formula of \cite{8}. In our case $J_0^{ on-shell}=0$, so our {formula} is a consequence of the general formula for entropy in WCFT iff
\be
L_0^{\it vac}-\frac{(J_0^{\it vac})^2}{2\kappa}=-\frac{\kappa}{8}\,.
\ee
One might intuitively expect that formula for the black hole entropy in WCFT should correctly reproduce the entropy of an extremal OTT black hole. The intuition behind it relies
on the resemblance of the algebra \eq{5.20}, \eq{5.21} with the Euclidean WCFT algebra and it is expected that the same derivation \cite{8} holds in our case.   
\section{Sugawara-Sommerfeld construction}
\setcounter{equation}{0}

It is well-known that one can construct the Virasoro algebra as a bilinear combination of the elements of the Kac-Moody algebra. Now, we apply this procedure, known as the Sugawara-Sommerfeld construction \cite{16}, to the algebra obtained in the previous section.

First we introduce auxiliary operators
\be
K_n=\frac{1}{2\kappa}\sum_iJ_iJ_{n-i},
\ee
which obey the following commutation relations
\bea
&&i\{K_m,J_n\}=-nJ_{m+n},\\
&&i\{K_m,K_n\}=(m-n)K_{m+n},\\
&&i\{K_m,L_n\}=(m-n)K_{m+n}.
\eea
Then, we define generators of the first Virasoro algebra as
\be
L^R_n=L_n-K_n,
\ee
which satisfy the commutation relations
\bea
&&i\{J_m,J_n\}=\kappa m\d_{m+n,0},\\
&&i\{J_m,L^R_n\}=0,\\
&&i\{L^R_m,L^R_n\}=(m-n)L^R_{m+n}.
\eea
The generators of the second Virasoro algebra are defined by
\be
L^L_n=-K_{-n}-in\alpha J_{-n}+\frac{c^L}{24}\d_{n,0}\,.
\ee
The generators $L^L_n$ and $L^R_n$ define the two commuting Virasoro algebras
\bea
&&i\{L^L_m,L^L_n\}=(m-n)L^L_{m+n}+\dis\frac{c^L}{12}m(m^2-1)\d_{m+n,0},\\
&&i\{L^L_m,L^R_n\}=0,\\
&&i\{L^R_m,L^R_n\}=(m-n)L^R_{m+n},
\eea
with central charges
\bea\lab{4.13}
c^L=12\kappa \a^2\,,\qquad c^R=0\,.
\eea

In theories with conformal symmetry it is well-known that the entropy can be reproduced by Cardy formula. Sugawara-Sommerfeld construction includes an arbitrary parameter $\a$ which value  is fixed by demanding that the Virasoro algebra satisfies certain canonical relation. We shall
fix it by requiring that Cardy formula
\be
S=2\pi\sqrt{\frac{L^L_0c^L}{6}}+2\pi\sqrt{\frac{L^R_0c^R}{6}},
\ee
reproduces entropy correctly.
For the orbifold, the values of the Virasoro zero modes are
\be
L^L_0=\frac{c^L}{24}\,,\qquad  L^R_0=\frac{r^2_0}{2\ell G},
\ee
which implies, in combination with \eq{4.13}, that Cardy formula gives the entropy
\be
S=\frac{\pi c^L}6=2\pi\k\a^2.
\ee
Consequently, we get
\be
\a^2=\frac{r_0}{2\ell}\,.
\ee

\section{Thermodynamics at extremality}
\setcounter{equation}{0}

There is an equivalent Cardy {formula} in which, instead of using background values of the Virasoro zero modes, one uses the temperature. Thus, the required additional piece of information is the temperature of the dual CFT, which will be derived from the black hole thermodynamics.

We start from  the first law of black hole thermodynamics
\be
\d E=T_H\d S+\Om\d J+\Phi_i\d q^i,
\ee
where $J$ is the angular momentum, $\Om$ is the  angular velocity, $q^i$ are additional conserved charges, and $\Phi_i$ are potentials conjugate to $q^i$. In case of the extremal black hole (for more details
on extremal black holes and the first law of thermodynamics, see \cite{18}), for which the Hawking temperature is zero, $T_H=0$, the first law implies that energy is a function of the conserved charges
\be
E_{Ext}=E_{Ext}(J_{Ext},q^i_{Ext}).
\ee
The corresponding generalized temperatures are defined by
\be
T_L=\frac{\pd S_{Ext}}{\pd J_{Ext}}\,,\qquad T_i=\frac{\pd S_{Ext}}{\pd q^i_{Ext}}\,,
\ee
where $T_L$  is called left moving temperature.

The entropy, energy and angular momentum  of the extremal OTT  black hole are given by:
\be
S_{Ext}=\frac{\pi r_0}{G}\,,\qquad E_{Ext}=\frac{r_0^2}{2\ell^2 G}\,,\qquad J_{Ext}=\frac{r_0^2}{2\ell G}\,.
\ee
From the variation of the entropy of extremal OTT
$$
\d S_{Ext}=\frac{\d J_{Ext}}{T_L},
$$
the left moving temperature is determined as
\be
T_L=\frac{r_0}{\pi\ell}.
\ee
In extremal case, the right moving temperature is zero
\be
T_R=0.
\ee
By demanding  that the alternative form of the Cardy formula,
$$
S_C=\frac{\pi^2}{{3}}T_Lc^L+\frac{\pi^2}{{3}}T_Rc^R,
$$
reproduces the entropy of the extremal OTT black hole, one concludes that the value of the $c^R$ is undetermined and the left central charge is two times bigger that the Brown-Henneaux central charge
\be
c^L=\frac{3\ell}{G}.
\ee
This can be used to fix the constant $\a$ appearing in the Sugawara-Sommerfeld construction. From equation \eq{4.13} and the previous formula, we derive
\be
\a =\frac12\,.
\ee

\section{Concluding remarks}

We investigated near horizon symmetry of both static and stationary OTT black hole in the quadratic PGT. In the static case, the corresponding asymptotic symmetry is trivial, whereas in the stationary case, the set of consistent asymptotic conditions
leads to a symmetry described by time reparametrization and the semi-direct sum of the centrally extended $u(1)$
Kac-Moody and the chiral Virasoro algebra. The improved asymptotic conditions that follow from the vanishing of torsion \eq{5.16} can be further strengthened, making thus time reparametrization a pure gauge.

The near horizon limit corresponds to deep infrared sector of the theory, which implies that only soft part of the charge survives. This means that the corresponding charges represent soft hair on the black hole horizon. Formula
 $$
 S=2\pi \sqrt{\frac12L^{\it on-shell}_0\kappa},
$$
 shows that there is an intimate relation between black hole entropy and soft hairs on the horizon, but to obtain a more precise information requires further examination of the subject.

Using the Sugawara-Sommerfeld construction, we build the Virasoro algebra as a bilinear combination of $u(1)$ Kac-Moody and chiral Virasoro algebra. The presence of conformal symmetry enables us to use the Cardy formula for entropy, which correctly reproduces the black hole entropy.

\section*{Acknowledgments}

We would like to thank Milutin Blagojevi\'c for careful reading of the manuscript and for
numerous 
useful suggestions. This work
was partially supported by the Serbian Science Foundation under Grant No.
171031.

\end{document}